\documentclass{article}

\PassOptionsToPackage{numbers, compress}{natbib}

\usepackage[final]{nips_2018}

\usepackage[utf8]{inputenc} 
\usepackage[T1]{fontenc}    
\usepackage{hyperref}       
\usepackage{url}            
\usepackage{booktabs}       
\usepackage{amsfonts}       
\usepackage{nicefrac}       
\usepackage{microtype}      
\usepackage{graphicx}
\usepackage{subfig}
\usepackage{amsmath}
\usepackage[table,xcdraw]{xcolor}
\setcitestyle{square}

\title{Automatic Diagnosis of Short-Duration 12-Lead ECG using a Deep Convolutional Network}

\author{
     Ant\^{o}nio H. Ribeiro\textsuperscript{1, 2, *}, Manoel Horta Ribeiro\textsuperscript{1}, Gabriela Paix\~{a}o\textsuperscript{1, 3}, Derick Oliveira\textsuperscript{1}, \\ \textbf{Paulo R. Gomes\textsuperscript{1, 3}, J\'{e}ssica A. Canazart\textsuperscript{1}, Milton Pifano\textsuperscript{1, 3}, Wagner Meira Jr.\textsuperscript{1}},\\ 
     \textbf{Thomas B. Sch\"{o}n\textsuperscript{2}, Antonio Luiz Ribeiro\textsuperscript{1, 3, \dag}}\\
     \textsuperscript{1} Universidade Federal de Minas Gerais, Brazil, \textsuperscript{2} Uppsala University, Sweden,\\
     \textsuperscript{3} Telehealth Center from Hospital das Cl\'{i}nicas da Universidade Federal de Minas Gerais, Brazil. \\
     \textsuperscript{*}\texttt{antonio-ribeiro@ufmg.br}, \textsuperscript{\dag}\texttt{tom@hc.ufmg.br}
}

\usepackage{xcolor} 

\begin{document}

\maketitle

\begin{abstract}
We present a model for predicting electrocardiogram (ECG) abnormalities in short-duration 12-lead ECG signals which outperformed medical doctors on the 4th year of their cardiology residency. Such exams can provide a full evaluation of heart activity and have not been studied in previous end-to-end machine learning papers. Using the database of a large telehealth network, we built a novel dataset with more than 2 million ECG tracings, orders of magnitude larger than those used in previous studies. Moreover, our dataset is more realistic, as it consist of 12-lead ECGs recorded during standard in-clinics exams. Using this data, we trained a residual neural network with 9 convolutional layers to map 7 to 10 second ECG signals to 6 classes of ECG abnormalities. Future work should extend these results to cover a large range of ECG abnormalities, which could improve the accessibility of this diagnostic tool and avoid wrong diagnosis from medical doctors.
\end{abstract}

\section{Introduction}
\label{sec:intro}

Cardiovascular diseases are the leading cause of death  worldwide~\cite{gbd2016causesofdeathcollaborators_global_2017} and the electrocardiogram (ECG) is a major diagnostic tool for this group of diseases. As ECGs transitioned from analogue to digital, automated computer analysis of standard 12-lead electrocardiograms gained importance in the process of medical diagnosis~\cite{willems_testing_1987}. However, limited performance of classical algorithms~\cite{willems_diagnostic_1991, shah_errors_2007} precludes its usage as a standalone diagnostic tool and relegates it to an ancillary role \cite{estes_computerized_2013}. 

End-to-end deep learning has recently achieved striking success in task such as image classification~\cite{krizhevsky_imagenet_2012} and speech recognition~\citep{hinton_deep_2012}, and there are great expectations about how this technology may improve health care and clinical practice~\cite{stead_clinical_2018, naylorc_prospects_2018, hinton_deep_2018}. So far, the most successful applications used a supervised learning setup to automate diagnosis from exams. Algorithms have achieved better performance than a human specialist on their routine workflow in diagnosing breast cancer~\cite{bejnordi_diagnostic_2017} and detecting certain eye conditions from eye scans~\cite{defauw_clinically_2018}. While efficient, training  deep neural networks using supervised learning algorithms introduces the need for large quantities of labeled data which, for medical applications, introduce several challenges, including those related to confidentiality and security of personal health information \cite{beck_protecting_2016}.

Standard, short-duration 12-lead ECG is the most commonly used complementary exam for the evaluation of the heart, being employed across all clinical settings: from the primary care centers to the intensive care units. While tracing cardiac monitors and long-term monitoring, as the Holter exam, provides information mostly about cardiac rhythm and repolarization, 12-lead ECG can provide a full evaluation of heart, including arrhythmias, conduction disturbances, acute coronary syndromes, cardiac chamber hypertrophy and enlargement and even the effects of drugs and electrolyte disturbances.

Although preliminary studies using deep learning methods~\cite{rajpurkar_cardiologistlevel_2017, shashikumar_detection_2018} achieve high accuracy in detecting specific abnormalities using single-lead heart monitors,  the use of such approaches for detecting the full range of diagnoses that can be obtained from a 12-lead, standard, ECG is still largely unexplored. 
A contributing factor for this is the shortage of full digital 12-lead ECG databases, since most ECG are still registered only on paper, archived as images, or in PDF format \cite{sassi_pdfecg_2017}. Most available databases comprise a few hundreds of tracings and no systematic annotation of the full list of ECG diagnosis~\cite{lyon_computational_2018}, limiting their usefulness as training datasets in a deep learning setting.

This lack of systematically annotated data is unfortunate, as training an accurate automatic method of ECG diagnosis from a standard 12-lead ECG would be greatly beneficial.The exams are performed in settings where, often, there are  no specialists to analyze and interpret the ECG tracings, such as in primary care centers and emergency units. Indeed, primary care and emergency department health professionals have limited diagnostic abilities in interpreting 12-lead ECGs~\cite{mant_accuracy_2007, veronese_emergency_2016}. This need is most acute in low and middle-income countries,  which are responsible for more than $75\%$ of deaths related to cardiovascular disease~\cite{worldhealthorganization_global_2014}, and where, often, the population does not have access to cardiologists with full expertise in ECG diagnosis.

The main contribution of this paper is to introduce a large-scale \textit{novel} dataset of labelled 12-lead ECGs exams and to train \textit{and validate} a residual neural network in this relevant setup. We consider 6 types of ECG abnormalities: 1st degree AV block (1dAVb), right bundle branch block (RBBB), left bundle branch block (LBBB), sinus bradycardia (SB), atrial fibrillation (AF) and sinus tachycardia (ST), considered representative of both rhythmic and morphologic ECG abnormalities.

\section{Related work}
\label{sec:related_work}
 
Classical ECG software, such as University of Glasgow's ECG analysis program~\cite{pwmacfarlane_university_2005}, extracts the main features of the ECG signal using signal processing techniques and use them as input for classifiers. 
A literature review of these methods is given by~\cite{jambukia_classification_2015}. In~\cite{rahhal_deep_2016} a different approach is taken, where the ECG features are learned using an unsupervised method and then used as input to a supervised learning method.

End-to-end deep learning presents an alternative to these two-step approaches, where the raw signal itself is used as input to the classifier. In~\cite{rubin_densely_2017, acharya_application_2017, rajpurkar_cardiologistlevel_2017} the authors make use of a convolutional neural network to classify ECG abnormalities. The network architecture used in~\cite{rajpurkar_cardiologistlevel_2017} is inspired by architectures used for image classification and we make use of a similar architecture in this paper. There are differences though, in particular when it comes to the number of layers, input type (we use 12-leads, while~\cite{rajpurkar_cardiologistlevel_2017} used a single lead) and the output layer used. Recurrent networks are used in~\cite{teijeiro_arrhythmia_2017, shashikumar_detection_2018}. A review of  recent machine learning techniques applied for ECG automatic diagnosis is given in~\cite{cantwell_rethinking_2018}. The aforementioned methods and others (such as random forest and bayesian methods) are compared and a more extensive list of references using those methods is provided. 

The major difference between this paper and other previous applications of end-to-end learning for ECG classification is on the dataset used for training and validating the model. The most common dataset used to design and evaluate ECG algorithms is the MIT-BIH arrhythmia database~\cite{goldberger_physiobank_2000}, which was used for training in~\cite{acharya_application_2017, rahhal_deep_2016} and for almost all algorithms in~\cite{jambukia_classification_2015}. This data set contain 30-minutes 2-leads ECG records from 47 unique patients. In~\cite{shashikumar_detection_2018} they used a dataset of 24-hour Holter ECG recordings collected from 2,850 patients at the University of Virginia (UVA) Heart Station. In~\cite{rajpurkar_cardiologistlevel_2017} they construct a new dataset containing labeled data of 64,121 ECG records from 29,163 unique patients who have used Zio Patch monitor. The PhysioNet 2017 Challenge, made available 12,186 entries dataset captured from the AliveCor ECG monitor containing between 9 and 61 seconds recordings~\cite{clifford_af_2017}. All these datasets were obtained from cardiac monitors and holter exams, where patients are usually monitored for several hours, and are restricted to one or two leads. Our dataset, on the other hand, consists of  short duration (7 to 10 seconds) 12-lead tracings obtained from in-clinics exams and is orders of magnitude larger than those used in previous studies, with well over 2 million entries.

\section{Data}
\label{sec:data}

The dataset used for \textit{training and validating} the model consists of 2,470,424 records from 1,676,384 different patients from 811 counties in the state of Minas Gerais/Brazil. The duration of the ECG recordings is between 7 and 10 seconds. The  data was obtained between 2010 and 2016 by a telediagnostic ECG system developed and maintained by the Telehealth Network of Minas Gerais (TNMG), led by the Telehealth Center from the Hospital das Cl\'{i}nicas of the Federal University of Minas Gerais. We developed an unsupervised methodology that classifies each ECG according to the free text in the expert report. We combine this result with two existing automatic ECG classifiers (Glasgow and Minnesota), using rules derived from expert knowledge and from the manual inspection of samples of the exams to obtain the ground truth. In several cases, we assigned the exams to be manually reviewed by medical students. This was done with around $34,000$ exams. This process is thoroughly explained in Appendix~\ref{sec:data_preprocessing}.

We split this dataset into training and validation set. The training set contains 98\% of the data. And the validation set consist of 2\% (approximately 50,000 exams) used for tuning the hyperparameters.

The dataset used for \textit{testing} the model consists of $953$ tracings from distinct patients. These were also obtained from TNMG's ECG system but using a more rigorous methodology for labelling the abnormalities. Two medical doctors with experience in electrocardiography have independently annotated the ECGs. When they agree, the common diagnosis is considered as ground truth. And, in case of \textit{any} disagreement, a third medical specialist, aware of the annotations from the other two, decided the diagnosis. Appendix~\ref{sec:ecg_abnormalities} contain information about the abnormalities that can be found in both the training/validation set and the test set.

\section{Model}
\label{sec:model}

\begin{figure}[t]
    \centering
	\includegraphics[width=0.6\textwidth]{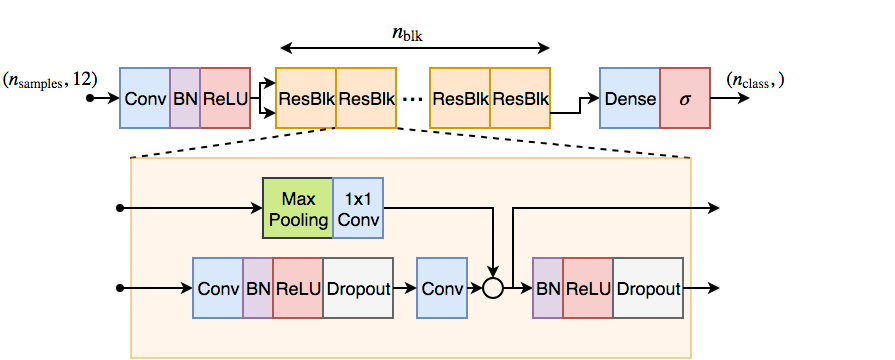}
    \caption{Unidimensional residual neural network used for ECG classification.}
    \label{fig:Resnet}
\end{figure}

We used a convolutional neural network similar to the residual network~\cite{he_deep_2015}, but adapted to unidimensional signals. This architecture allows deep neural networks to be efficiently trained by including skip connections. We have adopted the modification in the  residual block proposed in~\cite{he_identity_2016}, which place the skip connection in the position displayed in Figure~\ref{fig:Resnet}. A similar architecture has been successfully employed for arrhythmia detection from ECG signals in~\cite{rajpurkar_cardiologistlevel_2017} and the design choices we make in this section are, indeed, strongly influenced by~\cite{rajpurkar_cardiologistlevel_2017}. We should highlight that, despite using a significantly larger training dataset, we got the best validation results with an architecture with, roughly, one quarter the number of layers and parameters of the network employed in~\cite{rajpurkar_cardiologistlevel_2017}.

The network consists of a convolutional layer (\texttt{Conv})  followed by $n_{\text{blk}}=4$ residual blocks  with two convolutional layers per block.  The output of the last block is fed into a \texttt{Dense}  layer with sigmoid activation function ($\sigma$), which was used because the classes are not mutually exclusive (i.e. two or more classes may occur in the same exam). The output of each convolutional layer is rescaled using batch normalization,  \texttt{BN},~\cite{ioffe_batch_2015} and feed into a rectified linear activation unit, \texttt{ReLU}. \texttt{Dropout}~\cite{srivastava_dropout_2014} is applied after the non-linearity.

The convolutional layers have filter length 16, starting with 4096 samples and 64 filters for the first layer and residual block and increasing the number of filters by 64 every second residual block and subsampling by a factor of 4 every residual block.  \texttt{Max Pooling} and convolutional layers with filter length 1 (\texttt{1x1 Conv}) may be included in the skip connection to make the dimensions match the ones from signals in the main branch.

The loss function is the average cross-entropy  $\frac{1}{n_{\text{class}}} \sum_{i=1}^{n_{\text{class}}} y_i \log \hat{y}_i + (1 - y_i) \log (1 - \hat{y}_i)$ where $\hat{y}_i$ is the output of the sigmoid layer for the $i$-th class and $y_i$ is the corresponding observed value (0 or 1). The cost function (i.e. the sum of loss functions over the entire training set) is minimized using the Adam optimizer~\cite{kingma_adam_2014} with default parameters and learning rate $\text{lr} = 0.001$. The learning rate is reduced by a factor of 10 whenever the validation loss does not present any improvement for 7 consecutive epochs. The neural network weights are initialized as in~\cite{he_delving_2015} and the bias are initialized with zeros. The training runs for 50 epochs with the final model being  the one with best validation results during the optimization process.

\section{Results}

Table~\ref{tab:performance} shows the performance on the test set. We consider our model to have predicted the  abnormality when its output is above a threshold that is set manually for each of the classes. Each threshold was chosen to be approximately in the inflection point of the precision-recall curve (presented in  Appendix~\ref{sec:additional_experiments}). High performance measures were obtained for all ECG abnormalities, with F1 scores above $80\%$ and specificity indexes over $99\%$.

The same dataset was evaluated by two 4th year cardiology medical doctors, each one annotating half of the exams in the test set. Their average performance is given in the table for comparison and, considering the F1 score, the model outperforms them for 5 out of 6 abnormalities. 

\begin{table}[t]
    \centering
    \begin{tabular}{ccccccc||cc}
         & \multicolumn{2}{c}{Precision (PPV)} & \multicolumn{2}{c}{Recall (Sensitivity)} &  \multicolumn{2}{c}{Specificity} & \multicolumn{2}{c}{F1 Score}\\
         \cline{2-9}
         & model & doctor  & model & doctor & model & doctor & model & doctor\\
        \hline
        1dAVb & 0.923 & 0.905 & 0.727 & 0.679 & 0.998 & 0.998 & \textbf{0.813} & 0.776 \\
        RBBB & 0.878 & 0.868 & 1.000 & 0.971 & 0.995 & 0.994 & \textbf{0.935} & 0.917  \\
        LBBB  & 0.971 & 1.000 & 1.000 & 0.900 & 0.999 & 1.000 & \textbf{0.985} & 0.947 \\
        SB  & 0.792 & 0.833 & 0.864 & 0.938 & 0.995 & 0.996 & 0.826 & \textbf{0.882} \\
        AF & 0.846 & 0.769 & 0.846 & 0.769 & 0.998 & 0.996 & \textbf{0.846} & 0.769  \\
        ST & 0.870 & 0.938 & 0.952 & 0.833 & 0.993 & 0.998 & \textbf{0.909} & 0.882 \\
        \hline
    \end{tabular}
    \caption{Performance of our deep neural network model and 4th year cardiology resident medical doctors when evaluated on the test set. 
    (PPV = positive predictive value)}
    \label{tab:performance}
\end{table}

\section{Future Work}

The training data was collected from a general Brazilian population and, since the database is large, it contains even rare conditions with sufficient frequency so we can try to build models to predict them.  In future work we intend to extend the results to progressively larger classes of diagnosis. This process will happen gradually because: i) the dataset  preprocessing can be time consuming and demands a lot of work (Appendix~\ref{sec:data_preprocessing}); ii) generating validation data demand work hours of experienced medical doctors.

The Telehealth Center at the Hospital das Cl\'{i}nicas of the Federal University of Minas Gerais receives and assesses more than 2,000 digital ECGs per day. With the progressive improvements in the interface with the medical experts, the quality of this data should progressively increase, and it could be used in training, validating and testing future models. 

The Telehealth Center is currently serving more than 1000 remote locations in 5 Brazilians states and have the means to deploy and evaluate such automatic classification systems as a part of broader telehealth solutions, which could help to improve its capacity, making it possible to provide access of a broader population, with better quality reports.

\section{Conclusion}

These promising initial results point to end-to-end learning as a competitive alternative to classical automatic ECG classification methods. The development of such technologies may yield high-accuracy automatic ECG classification systems that could save clinicians considerable time and prevent wrong diagnosis. Millions of 12-lead ECGs are performed every year, many times in places where there is a shortage of qualified medical doctors to interpret them. An accurate classification system could help detecting wrong diagnosis and improve the access of patients from deprived and remote locations to this essential diagnostic tool of cardiovascular diseases.

\subsubsection*{Acknowledgments}

This research was partly supported by the Brazilian Research Agencies CNPq, CAPES, and FAPEMIG, by projects InWeb, MASWeb, EUBra-BIGSEA, INCT-Cyber and Atmosphere, and by the \emph{Wallenberg AI, Autonomous Systems and Software Program (WASP)} funded by Knut and Alice Wallenberg Foundation. We also thank NVIDIA for awarding our project with a Titan V GPU.

\bibliographystyle{ieeetr}
\bibliography{refs}

\appendix
\section{Training data preprocessing}
\label{sec:data_preprocessing}

In this appendix, we detail the preprocessing  of the data used for training and validating the model. The exams were analyzed by doctors during routine workflow and are subject to medical errors, moreover there might be errors associated with the semi-supervised methodology used to extract the diagnoses.
Hence, we combine the expert annotation with well established automatic classifiers to improve the quality of the dataset. Given i) the exams in the database; ii) the diagnoses given by the Glasgow and Minnesota automatic classifiers (\textit{automatic diagnosis}); and, iii) the diagnoses extracted from the expert free text associated with the exams using the unsupervised methodology (\textit{medical diagnosis}), the following procedure is used for obtaining the ground truth annotation:

\begin{enumerate}

\item We:
\begin{enumerate}
    \item \textit{Accept a diagnosis} (consider an abnormality to be present) if both the expert \emph{and} either the Glasgow or the Minnesota automatic classifiers indicated the same abnormality. 
    \item \textit{Reject a diagnosis} (consider an abnormality to be absent) if only one classifier indicates the abnormality in disagreement with both the doctor and the other automatic classifier. 
\end{enumerate}
After this initial step diagnoses there are two scenarios where we still need to accept or reject diagnoses. They are: 
i) both classifiers indicate the abnormality but the expert doesn't; or 
ii) only the expert indicates the abnormality but no classifier does.

\item We used some rules \textit{to reject some of the remaining diagnoses}:

\begin{enumerate}

    \item Diagnoses of ST where the heart rate was below $100$ ($8376$ medical diagnoses and $2$ automatic diagnoses)  were \textit{rejected}. 
    \item Diagnoses of SB where the heart rate was above $50$ ($7361$ medical diagnoses and $16427$ automatic diagnosis) were \textit{rejected}. 
    \item Diagnoses of LBBB or RBBB where the duration of the QRS interval was below $115$ ms ($9313$ medical diagnoses for RBBB and $8260$ for LBBB)  were \textit{rejected}. 
    \item Diagnoses of 1dAVb where the duration of the PR interval was below $190$ ms ($3987$ automatic diagnoses) were \textit{rejected}.
\end{enumerate}

\item Then, using the sensitivity analysis of $100$ manually reviewed exams per abnormality, we came up with the following rules \textit{to accept some diagnoses remaining}:
\begin{enumerate}
    \item For RBBB, d1AVb, SB and ST we \textit{accepted} all medical diagnoses. $26033$, $13645$, $12200$ and $14604$ diagnoses were \textit{accepted} in such fashion, respectively
    \item For FA, we required not only that the exam was classified by the doctors as true but also that the standard deviation of NN intervals was higher than $646$. $14604$ diagnoses were \textit{accepted} using this rule.
\end{enumerate}
According to the sensitivity analysis  the number of false positives that would be introduced  by this procedure was smaller than $3\%$ of the total number of exams.

\item After this process, we were still left with a approximately $34000$ exams whose diagnoses had not been accepted or rejected. These were \textit{manually reviewed} by medical students using the Telehealth ECG diagnostic system. The process of manually reviewing these $34000$ ECGs took several months.
\end{enumerate}

\section{ECG abnormalities}
\label{sec:ecg_abnormalities}

\begin{table}[htpb]
  \centering
  \caption{Prevalence of each abnormality in the train/validation set and in the test set. It contains both the percentage \% and the absolute number of patients (in parentheses).}
  \begin{tabular}{cp{4cm}p{2.5cm}p{2.5cm}}
    \hline
    Abbrev. & Description & Prevalence (Train+Val) & Prevalence (Test)\\
    \hline
    1dAVb & 1st degree AV block &  1.5 \% (36,324) &  3.5 \% (33)\\
    RBBB & Right bundle branch block  &  2.6\% (64,319) &  3.8 \% (36)\\
    LBBB & Left bundle branch block &  1.5\% (37,326) &  3.5 \% (33)\\
    SB & Sinus bradycardia &  1.6\% (38,837) &  2.3 \% (22)\\
    AF & Atrial fibrilation & 1.7\% (42,133) &  1.4 \% (13)\\
    ST & Sinus tachycardia  & 2.3\% (56,186) &  4.4 \% (42) \\
    \hline
  \end{tabular}
  \label{tab:diagnosis}
\end{table}

\begin{figure}[htpb]
    \centering
    \includegraphics[width=0.8\textwidth]{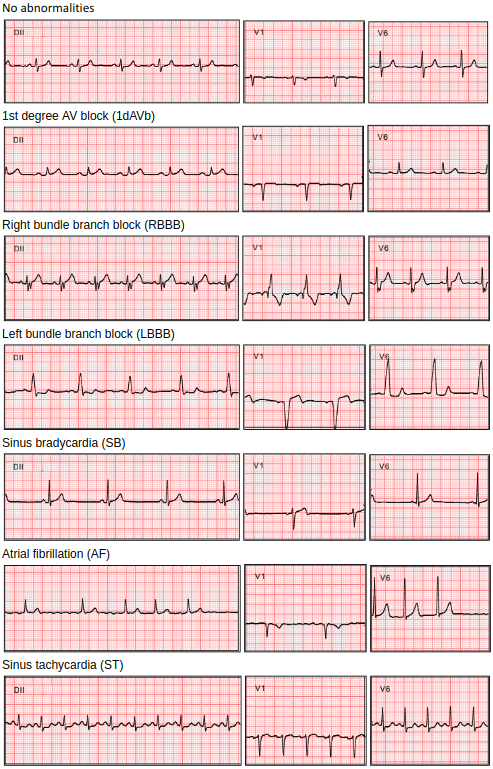}
    \caption{A list of all the abnormalities the model classifies. We show only 3 representative leads (DII, V1 and V6).}
    \label{fig:}
\end{figure}

\newpage

\section{Additional experiments}
\label{sec:additional_experiments}

\begin{figure}[htpb]
    \centering
    \subfloat[][1dAVb (average precision = 0.91)]{\includegraphics[width=0.5\textwidth]{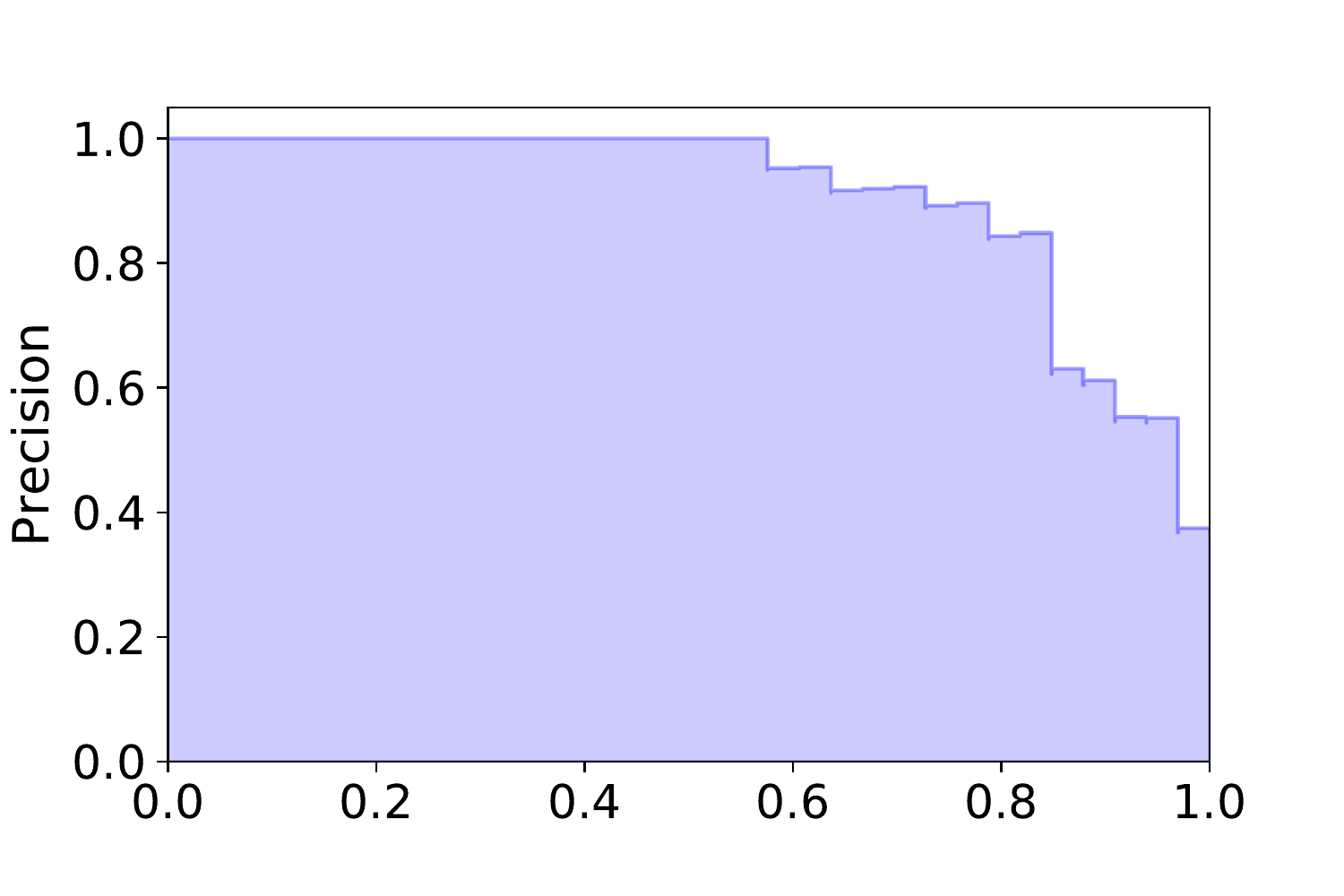}}
    \subfloat[][RBBB (average precision = 0.94)]{\includegraphics[width=0.5\textwidth]{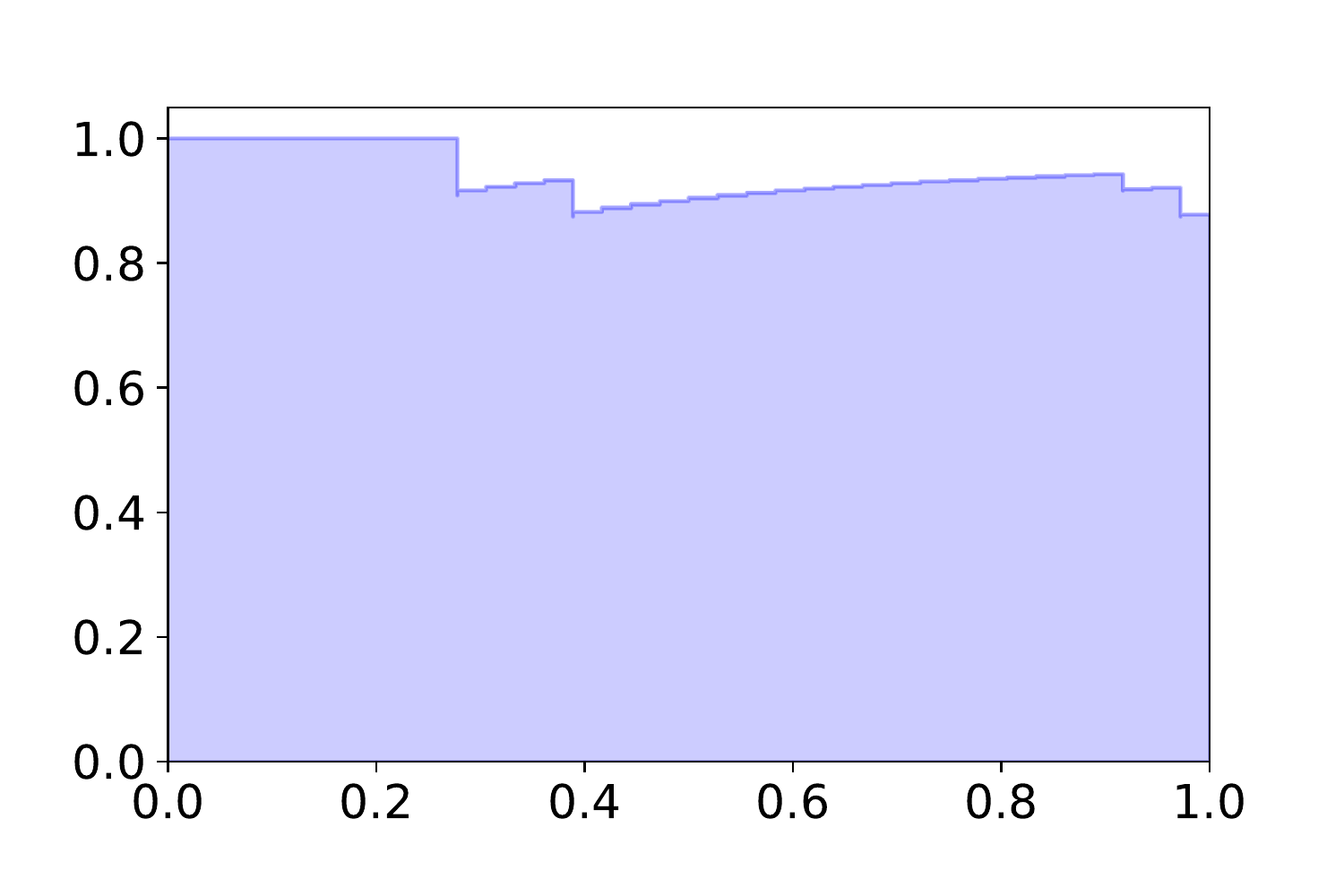}}\\
    \subfloat[][LBBB (average precision = 1.00)]{\includegraphics[width=0.5\textwidth]{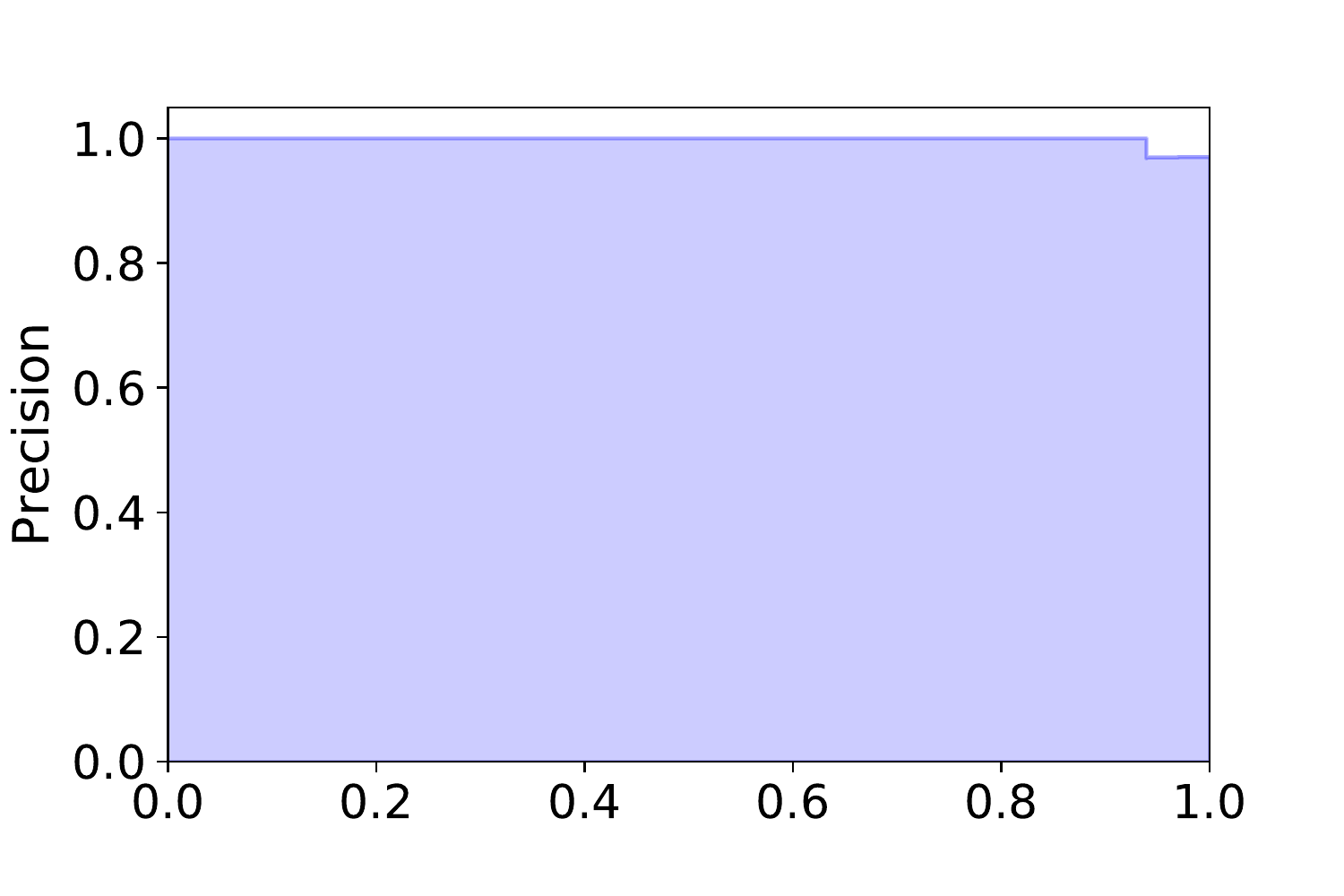}}
    \subfloat[][SB (average precision = 0.87)]{\includegraphics[width=0.5\textwidth]{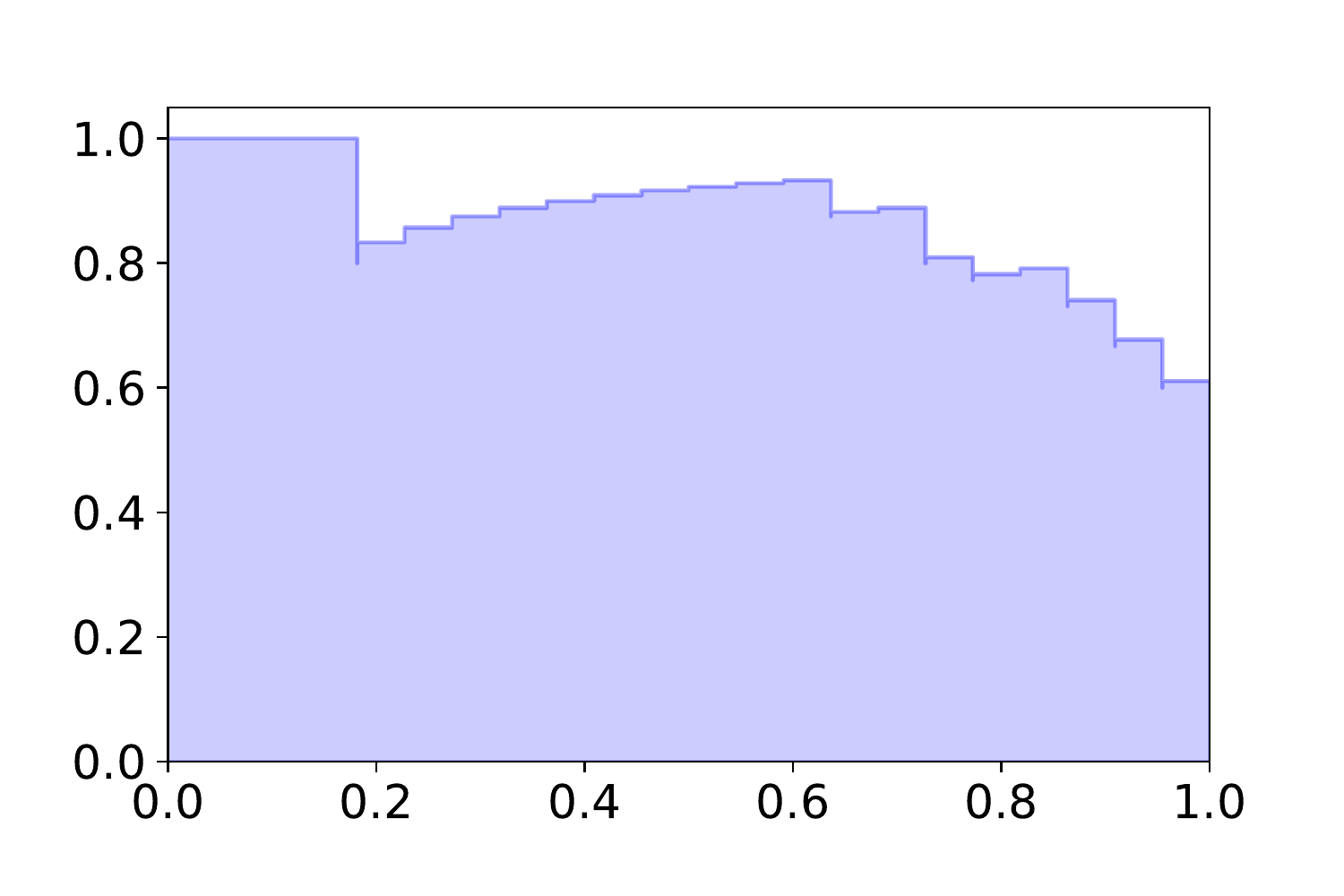}}\\
    \subfloat[][AF (average precision = 0.90)]{\includegraphics[width=0.5\textwidth]{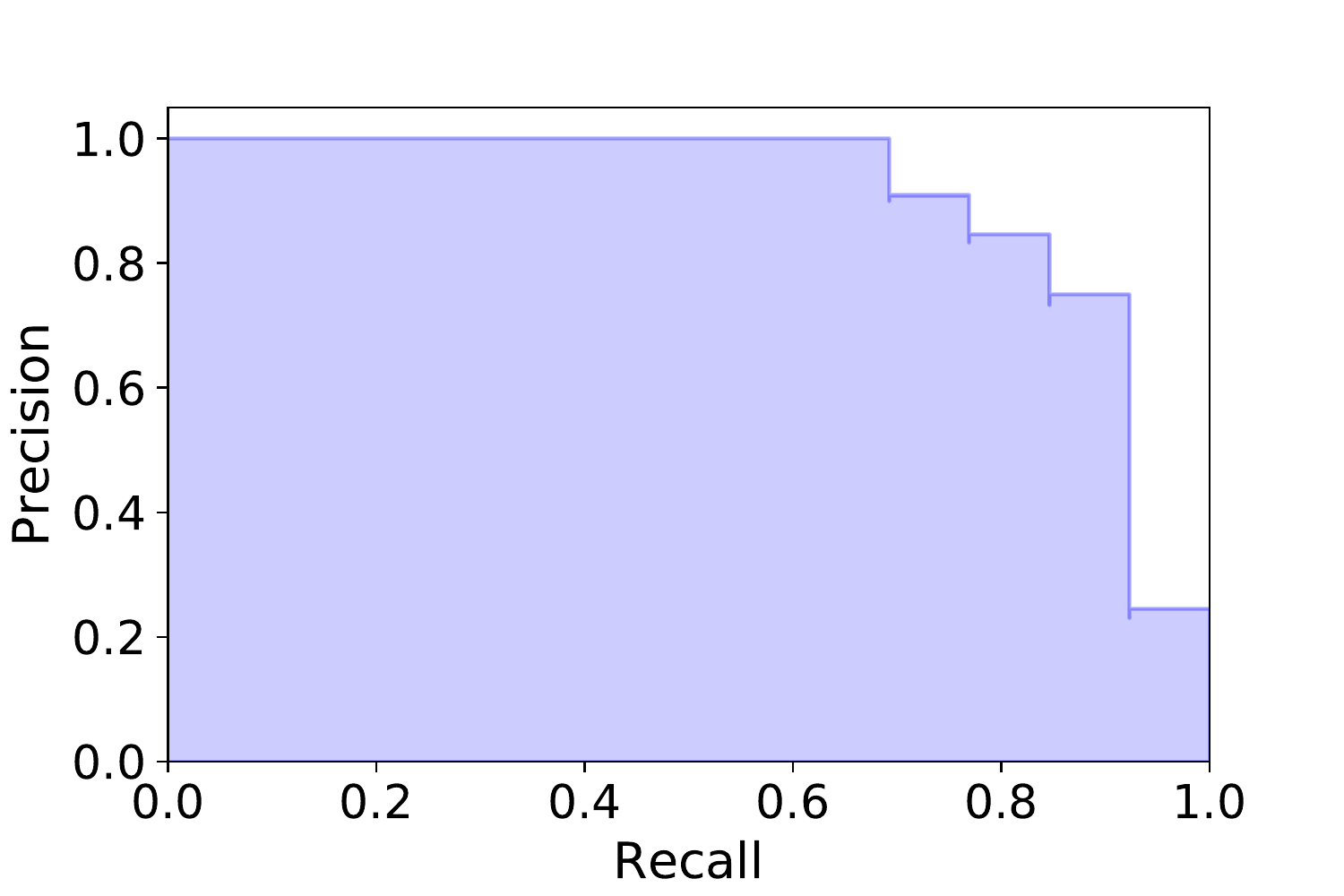}}
    \subfloat[][ST (average precision = 0.96)]{\includegraphics[width=0.5\textwidth]{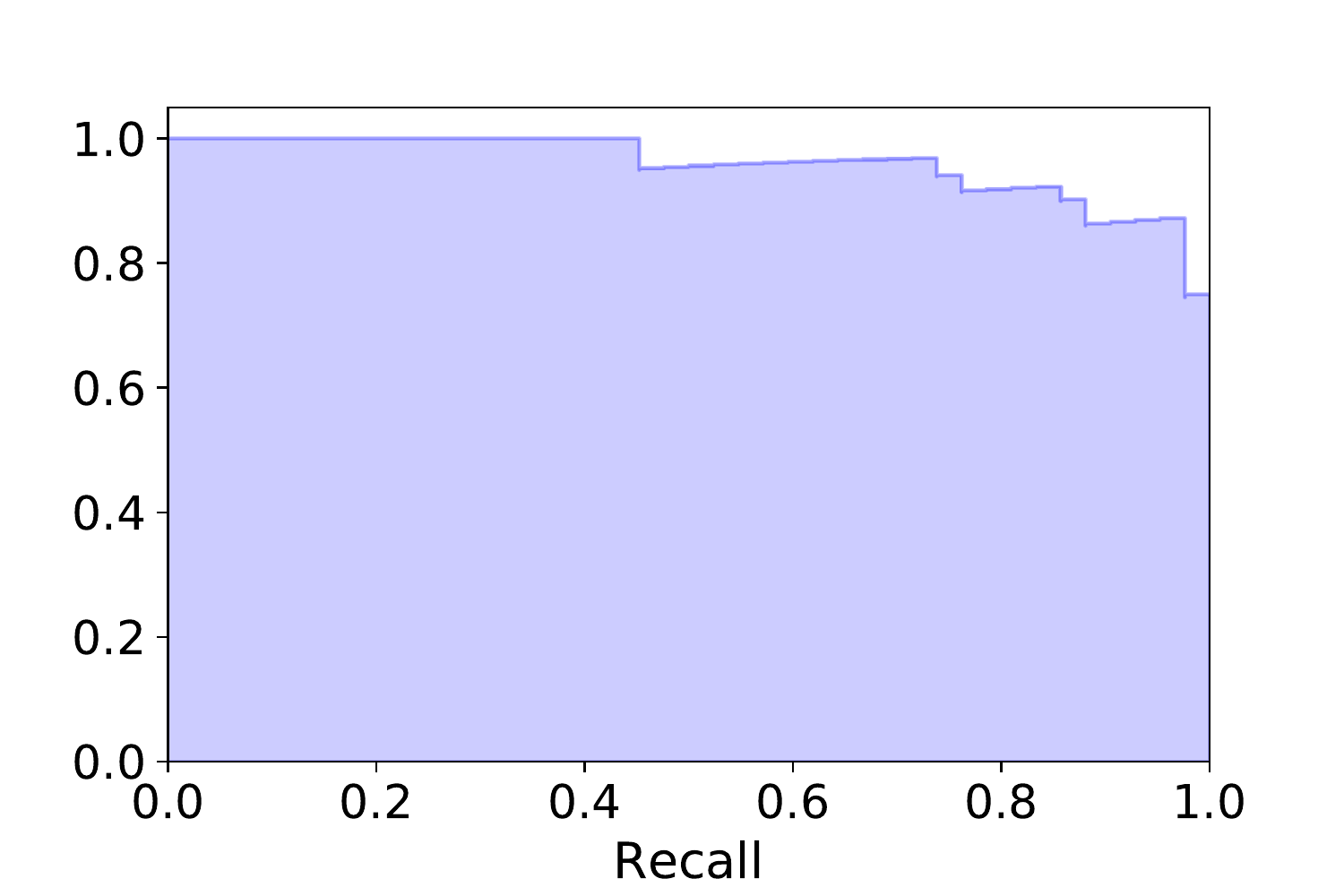}}
    \caption{Precision-recall  curve for our prediction model in the test set with regard to each ECG abnormalities. The average precision (which is approximated by the area under the precision-recall curve) is displayed in the captions.}
    \label{fig:precision_recall}
\end{figure}

In Figure~\ref{fig:precision_recall} we show the precision-recall curve for our model. This is a useful graphical representation to assess the success of a prediction model when, as in our case, the classes are imbalanced. The thresholds we used to generate Table~\ref{tab:performance} were chosen trying to get the inflection point of these curves. And, for these same thresholds, Table~\ref{tab:confusion_matrices} show the neural network confusion matrix for each of the classes.

\begin{table}[ht]
\centering
\begin{tabular}{ccccccc}
             & \multicolumn{2}{c}{Predicted Class}                                           &                       &              & \multicolumn{2}{c}{Predicted Class}                      \\
Actual Class & 1dAVb                      & \multicolumn{1}{c|}{Not 1dAVb}                   & \multicolumn{1}{c|}{} & Actual Class & RBBB                       & Not RBBB                    \\
1dAVb        & \cellcolor[HTML]{9AFF99}\textbf{24}& \multicolumn{1}{c|}{\cellcolor[HTML]{CDDDAC}9}   & \multicolumn{1}{c|}{} & RBBB         & \cellcolor[HTML]{9AFF99}\textbf{36} & \cellcolor[HTML]{CDDDAC}0   \\
Not 1dAVb    & \cellcolor[HTML]{CDDDAC}2  & \multicolumn{1}{c|}{\cellcolor[HTML]{9AFF99}\textbf{918}} & \multicolumn{1}{c|}{} & Not RBBB     & \cellcolor[HTML]{CDDDAC}5  & \cellcolor[HTML]{9AFF99}\textbf{912} \\ \cline{1-3} \cline{5-7} \\
Actual Class & LBBB                       & \multicolumn{1}{c|}{Not LBBB}                    & \multicolumn{1}{c|}{} & Actual Class & SB                         & Not SB                      \\
LBBB         & \cellcolor[HTML]{9AFF99}\textbf{33} & \multicolumn{1}{c|}{\cellcolor[HTML]{CDDDAC}0}   & \multicolumn{1}{c|}{} & SB           & \cellcolor[HTML]{9AFF99}\textbf{19} & \cellcolor[HTML]{CDDDAC}3   \\
Not LBBB     & \cellcolor[HTML]{CDDDAC}1  & \multicolumn{1}{c|}{\cellcolor[HTML]{9AFF99}\textbf{919}} & \multicolumn{1}{c|}{} & Not SB       & \cellcolor[HTML]{CDDDAC}5  & \cellcolor[HTML]{9AFF99}\textbf{926} \\ \cline{1-3} \cline{5-7} \\
Actual Class & AF                         & \multicolumn{1}{c|}{Not AF}                      & \multicolumn{1}{c|}{} & Actual Class & ST                         & Not ST                      \\
AF           & \cellcolor[HTML]{9AFF99}\textbf{11} & \multicolumn{1}{c|}{\cellcolor[HTML]{CDDDAC}2}   & \multicolumn{1}{c|}{} & ST           & \cellcolor[HTML]{9AFF99}\textbf{40} & \cellcolor[HTML]{CDDDAC}2   \\
Not AF       & \cellcolor[HTML]{CDDDAC}2  & \multicolumn{1}{c|}{\cellcolor[HTML]{9AFF99}\textbf{938}} & \multicolumn{1}{c|}{} & Not ST       & \cellcolor[HTML]{CDDDAC}6  & \cellcolor[HTML]{9AFF99}\textbf{905}\\ \cline{1-3} \cline{5-7} \\
\end{tabular}
\caption{Confusion matrices for the neural network.}
    \label{tab:confusion_matrices}
\end{table}

\end{document}